\documentclass[a4paper,11pt]{article}
\usepackage{pos}
\usepackage{slashed}
\title{Probing the structure of $\chi_{c1}(3872)$: \\Heavy quark symmetries at work}
\ShortTitle{Heavy quark symmetries at work}
\author*[a,b]{Giuseppe Roselli}

\affiliation[a]{Dipartimento Interuniversitario di Fisica ``M.~Merlin'', Universit\`a degli Studi di Bari, Via Orabona 4, 70125 Bari, Italy}

\affiliation[b]{INFN, Sezione di Bari, Via Orabona 4, 70125 Bari, Italy}

\emailAdd{giuseppe.roselli@ba.infn.it}

\abstract{More than two decades have elapsed since the discovery of $\chi_{c1}(3872)$. For this meson,
previously denoted as $X(3872)$, an impressive amount of theoretical and
experimental studies has been devoted concerning its properties, decays and production
mechanisms. Despite the extensive work, a full understanding of the nature of
$\chi_{c1}(3872)$ is missing.
I describe a theoretical framework based on the heavy
quark large mass limit to analyze the radiative decays of heavy quarkonia, in
particular the electric dipole transitions of $\chi_{c1}(2P)$ to $S$-wave
charmonia. The results favorably compare to recent LHCb collaboration  measurements for
$\chi_{c1}(3872)$, if this meson is
identified with $\chi_{c1}(2P)$.
}

\FullConference{Fourth Italian Workshop on Physics at High Intensity (WIFAI2025)\\
11–14 November 2025\\
Bari, Italy\\}

\begin{document}
\maketitle
\section{Introduction}

In recent years, several states with properties that challenge their classification as quark-antiquark mesons or three quark baryons have been observed. Such states, referred to as exotic states, provide insights about the nonperturbative regime of QCD. Here, I describe a study \cite{Colangelo:2025uhs} on  $X(3872)$, also called $\chi_{c1}(3872)$ \cite{ParticleDataGroup:2024cfk}, one of the first exotic candidates, detected by the Belle collaboration \cite{Belle:2003nnu} as a peak in the $J/\psi \pi^+ \pi^-$ invariant mass distribution in the $B \to K \pi^+ \pi^- J/\psi$ decays. Several other experiments \cite{CDF:2003cab,BaBar:2004iez,D0:2004zmu,LHCb:2011zzp,BESIII:2013fnz,CMS:2013fpt,ATLAS:2016kwu,LHCb:2013kgk, LHCb:2020xds} allowed the determination of the quantum numbers $J^{PC} = 1^{++}$ and hint the assignment $I^G=0^+$ \cite{ParticleDataGroup:2024cfk}. An ordinary charmonium state with these quantum numbers is $\chi_{c1}(2P)$, still unobserved.
$\chi_{c1}(3872)$ presents various puzzling features.

The ratio ${\cal B}(\chi_{c1}(3872) \to J/\psi\, 3\pi) / {\cal B}(\chi_{c1}(3872) \to J/\psi\, 2\pi)$ is approximately equal to unity \cite{Belle:2005lfc,BaBar:2010wfc,BESIII:2019qvy}. Assuming a charmonium interpretation ($I=0$), this implies a large violation of the isospin symmetry, since the intermediate state would be $\omega$ ($I=0$) and $\rho^0$ ($I=1$), respectively, for the two decay channels.

To overcome this difficulty, non-conventional interpretations have been proposed~\cite{Brambilla:2019esw,Meyer:2015eta}. In particular, the mass of $\chi_{c1}(3872)$, $m = 3871.64 \pm 0.06$ MeV, is close to the $D^0\bar{D}^{*0}$ threshold, $M_{D^0} + M_{\bar{D}^{*0}} = 3871.69 \pm 0.07$ MeV, suggesting a molecular interpretation~\cite{Braaten:2003he,Swanson:2003tb,Wong:2003xk,Hanhart:2007yq}, with $\chi_{c1}(3872)$ being a loosely bound $D^0\bar{D}^{*0}$ system. Other possibilities include tetraquarks or mixing of different components~\cite{Germani:2025mos}.

An observable sensitive to the structure of the $\chi_{c1}(3872)$ is the ratio \cite{Swanson:2004pp}

\begin{equation}
\label{RadiativeR}
\mathcal{R}(\chi_{c1}(3872)) = 
\frac{{\cal B}(\chi_{c1}(3872) \to \psi(2S)\gamma)}
{{\cal B}(\chi_{c1}(3872) \to J/\psi\, \gamma)} \, .
\end{equation}

\noindent
The most recent experimental determination of this ratio has been provided by the LHCb collaboration: $\mathcal{R} = 1.67 \pm 0.21 \pm 0.12 \pm 0.04$ \cite{LHCb:2024tpv}.

In Ref. \cite{Colangelo:2025uhs}, radiative decays of charmonia and bottomonia have been analyzed. In particular, in a theoretical approach based on the heavy quark spin symmetry,  under the assumption that $\chi_{c1}(3872)$ is identified as $\chi_{c1}(2P)$, the ratio $\mathcal{R}$ in Eq.\eqref{RadiativeR} is found in good agreement with the LHCb result.

\section{Heavy quark symmetries in QCD}
Systems containing heavy quarks can be suitably studied exploiting heavy-quark symmetries.
In hadronic systems containing a single heavy quark (HQ), the large HQ mass allows simplifications formalized in the Heavy Quark Effective Theory (HQET) \cite{Neubert:1993mb}.
The HQ mass limit, $m_Q \gg \Lambda_{\mathrm{QCD}}$, implies that the HQ decouples from the rest of the hadron, behaving as a static source of color field; the light degrees of  freedom of the system are insensitive to the spin and to the flavour of HQ. This implies new symmetries: invariance under HQ spin rotation and HQ flavour symmetry.
The HQET Lagrangian can be obtained from the QCD Lagrangian defining the field $h_v(x)=e^{im_Q v \cdot x}P_+Q(x)$ where $Q$ is the HQ field in QCD, $P_+=\displaystyle{1 + \slashed{v} \over 2}$ and $v$ is the HQ velocity. The HQ momentum can be written as $p=m_Q v+k$, where $k$ is a residual momentum of ${\cal O}(\Lambda_{QCD})$.
In terms of $h_v$ the HQET Lagrangian reads $\mathcal{L}_{HQET}={\bar h}_v i \, v \cdot D h_v$, $D$ being the QCD covariant derivative. Symmetry breaking terms arise, including subleading operators suppressed by powers of $k/m_Q$. At ${\cal O}(1/m_Q)$ two operators appear
\begin{equation}
{\cal L}^{(1)}={1 \over 2 m_Q} {\bar h}_v (i \slashed{D}_\perp)^2 h_v +{1 \over 2 m_Q}{\bar h}_v {g_s \sigma_{\alpha \beta}
G^{\alpha \beta} \over 2 } h_v 
\label{lag1m} \; ,
\end{equation}
they represent the  kinetic energy of the heavy quark
due to its residual momentum $k$, and the
chromomagnetic coupling of the heavy quark spin to the gluon field, respectively.

When we are dealing with quarkonia systems, considering gluon exchanges between two HQ with the same velocity, infrared (IR) divergences arise \cite{Thacker:1990bm}. To regulate this behavior it is necessary to include the first term in Eq. \eqref{lag1m} in the leading order Lagrangian. Thus, only HQ spin symmetry survives, implying that hadrons can be grouped in multiplets that differ only for the orientation of the HQ spin and are degenerate in the HQ limit.

The multiplets describing $S$- and $P$-wave states, which correspond to $L=0$ and $L=1$, respectively, can be written as:
\begin{itemize}
\item $L=1$ multiplet:
\begin{equation}
J^{\mu} =
\frac{1+\slashed{v}}{2}\Bigg[
H_2^{\mu \alpha} \gamma_\alpha
+ \frac{1}{\sqrt{2}}
\epsilon^{\mu \alpha \beta \gamma}
v_\alpha \gamma_\beta H_{1\gamma}
+ \frac{1}{\sqrt{3}}
(\gamma^{\mu} - v^{\mu}) H_0
+ K_1^{\mu}\gamma_5
\Bigg]\frac{1-\slashed{v}}{2} \, ,
\label{Pwave}
\end{equation}
\item $L=0$ multiplet:
\begin{equation}
J =
\frac{1+\slashed{v}}{2}
\left[ H_1^\mu \gamma_\mu - H_0 \gamma_5 \right]
\frac{1-\slashed{v}}{2} \, .
\label{Swave}
\end{equation}
\end{itemize}
We focus on radiative decays and construct effective interaction Lagrangians with photons in terms of the HQ spin multiplets.
\section{Radiative decays of heavy quarkonia}
Let us consider electric dipole transitions with $\Delta L = 1$, in particular transitions from $P$ to $S$-wave states. The most general Lagrangian invariant under $C$, $P$, $T$ reads \cite{Casalbuoni:1993cx, DeFazio:2008xq}:
\begin{equation} {\cal L}_{nP \leftrightarrow mS}=\delta^{nPmS}_Q {\rm Tr}
\left[{\bar J}(mS) J_\mu(nP) \right] v_\nu F^{\mu \nu} + \rm{h.c.}
\,.\label{lagPS} 
\end{equation}
$F^{\mu \nu}$ is the electromagnetic field
strength tensor and $\delta^{nPmS}_Q$ is a coupling governing all transitions from the $nP$ multiplet to the $mS$ one.
From this Lagrangian, one can derive expressions for the decay widths:
\begin{equation}
\begin{aligned}
\Gamma(n^3P_J \to m^3S_1 \gamma) &=
\frac{(\delta_Q^{nPmS})^2}{3\pi}\,
k_\gamma^3\,\frac{M_{S_1}}{M_{P_J}}
\\
\Gamma(m^3S_1 \to n^3P_J \gamma) &=
(2J+1)\frac{(\delta_Q^{nPmS})^2}{9\pi}\,
k_\gamma^3\,\frac{M_{P_J}}{M_{S_1}}
 \\
\Gamma(n^1P_1 \to m^1S_0 \gamma) &=
\frac{(\delta_Q^{nPmS})^2}{3\pi}\,
k_\gamma^3\,\frac{M_{S_0}}{M_{P_1}}
 \\
\Gamma(m^1S_0 \to n^1P_1 \gamma) &=
\frac{(\delta_Q^{nPmS})^2}{\pi}\,
k_\gamma^3\,\frac{M_{P_1}}{M_{S_0}}
\, .
\end{aligned}
\label{decays}
\end{equation}
\section{Numerical results: Radiative decays}
\begin{figure}[t]
\begin{center}
\includegraphics[width = 0.85\textwidth]{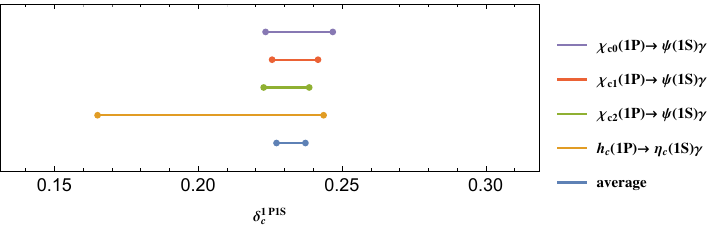}
\caption{\baselineskip 10pt  \small  Coupling $\delta_c^{1P1S}$ obtained for the modes listed in the legenda. The average value is also displayed.}\label{fig:deltac1P1S}
\end{center}
\end{figure}

To verify the validity of the approach, we consider well measured decay channels, starting from the $P$-wave multiplet with $n=1$ ($\chi_{c0}(1P),\,\chi_{c1}(1P),\,\chi_{c2}(1P),\,h_c(1P)$) to $S$-wave states with $n=1$ $(\eta_c(1S),\,J/\psi)$.
Using Eq.\eqref{decays}, we extract $\delta^{1P1S}$. In Fig.~\ref{fig:deltac1P1S} we show that the four values extracted are mutually compatible, as expected in the HQ limit.

In order to derive predictions for \(\chi_{c1}(3872)\), if it is the \(\chi_{c1}(2P)\) state belonging to the multiplet (\(\chi_{c0}(2P), \chi_{c1}(2P), \chi_{c2}(2P), h_c(2P)\)), we consider, as a starting point, the corresponding beauty sector. Since the total widths are not yet known in this sector, we consider instead:
$${\tilde \delta}_b^{nPmS}(P_b)=\displaystyle\frac{\delta_b^{nPmS}}{[\Gamma_{\rm tot}(P_b)]^{1/2}}\ ,$$
where \(P_b\) denotes a state in an \(nP\) multiplet.
We can extract the $\tilde{\delta}$ for the transitions from the multiplet 
$(\chi_{b0}(2P),\,\chi_{b1}(2P),\,\chi_{b2}(2P),\,h_b(2P))$
to the two doublets $(\eta_b(1S),\,\Upsilon(1S))$ and $(\eta_b(2S),\,\Upsilon(2S))$ exploiting the results for the branching ratios in PDG \cite{ParticleDataGroup:2024cfk}:
\begin{align}
{\tilde \delta}_b^{2P1S} (\chi_{b0}(2P)) &= (0.31 \pm 0.07) & {\tilde \delta}_b^{2P2S} (\chi_{b0}(2P)) &= (3.9 \pm 0.4) \nonumber \\
{\tilde \delta}_b^{2P1S} (\chi_{b1}(2P)) &= (1.51 \pm 0.08) & {\tilde \delta}_b^{2P2S} (\chi_{b1}(2P)) &= (12.0 \pm 0.6) \nonumber \\
{\tilde \delta}_b^{2P1S} (\chi_{b2}(2P)) &= (1.20 \pm 0.07) & {\tilde \delta}_b^{2P2S} (\chi_{b2}(2P)) &= (7.8 \pm 0.5) \\
{\tilde \delta}_b^{2P1S} (h_b(2P)) &= (2.01 \pm 0.23) & {\tilde \delta}_b^{2P2S} (h_b(2P)) &= (16.5 \pm 2.2) \, .\nonumber 
\end{align}

\noindent Defining the ratio $R_Q^\delta = \frac{\delta_Q^{2P2S}}{\delta_Q^{2P1S}}$ we have:
\begin{align}
R_b^\delta(\chi_{b0}(2P)) &= 13 \pm 4 
&\qquad
R_b^\delta(\chi_{b1}(2P)) &= 8 \pm 1 \nonumber\\
R_b^\delta(\chi_{b2}(2P)) &= 7 \pm 1
&\qquad
R_b^\delta(h_b(2P)) &= 8 \pm 2 \, ,
\end{align}
\noindent
where we have indicated the decaying particle whose radiative transitions are used to determine the value of $R_b^\delta$. Averaging these results we obtain
\begin{equation}
R_b^\delta = 9.0 \pm 0.7 \label{Rb} \,\,.
\end{equation}
Since we are considering a quarkonium system, we expect that HQ flavour symmetry is broken, which implies that $\delta_c^{nPmS} \neq \delta_b^{nPmS}$. However, it is reasonable to expect $\delta$ scales as $\delta_Q \simeq 1/m_Q$. 
This means $R_b^\delta \simeq R_c^\delta$, allowing us to derive several predictions.

We now express $\mathcal{R}$ in Eq.\eqref{RadiativeR} in terms of $R_c^\delta$
\begin{equation}
{\cal R}(\chi_{c1}(3872))=
\displaystyle\frac{{\cal B}(\chi_{c1}(3872) \to \psi(2S) \gamma)}
{{\cal B}(\chi_{c1}(3872) \to J/\psi \gamma)}
=(PS)\,\left[R_c^\delta (\chi_{c1}(3872))\right]^2 \,\, ,
\end{equation}
where (PS) is a phase space factor. Assuming $R_c^\delta \approx R_b^\delta$ we predict
\begin{equation}
{\cal R}(\chi_{c1}(3872))=1.7 \pm 0.3 \,\,\,. \label{r1}
\end{equation}
This agrees with the LHCb measurement \cite{LHCb:2024tpv}
\begin{equation}
{\cal R}(\chi_{c1}(3872))_{\rm LHCb}=1.67 \pm 0.21 \pm 0.12 \pm 0.04  \,\,\,.
\label{lhcb}
\end{equation}

\noindent
Moreover, the result \eqref{r1} together with the experimental value for 
${\cal B}(\chi_{c1}(3872) \to J/\psi \gamma)$ gives:
\begin{equation}
{\cal B}(\chi_{c1}(3872) \to \psi(2S) \gamma)
=(1.3 \pm 0.5)\times 10^{-2}\,\,.
\end{equation}
\section{Conclusions}
I have described the consequences of identifying $\chi_{c1}(3872)$ with $\chi_{c1}(2P)$, exploiting HQ spin symmetry \cite{Colangelo:2025uhs}. In particular, I derived ${\cal R}(\chi_{c1}(3872))$ obtaining a result in good agreement with the latest experimental determination.

The main strength of the analysis lies in its largely model-independent nature, as it is based on QCD in the heavy-quark limit, without free parameters to be tuned. Several other predictions have also been obtained. Their experimental verification would provide further support for both the method and the identification of $\chi_{c1}(3872)$ as $\chi_{c1}(2P)$ \cite{Colangelo:2025uhs}.
However, several issues remain unresolved, in particular understanding isospin breaking in $\chi_{c1}$ decays.

\section*{Acknowledgments}
This study was carried out within the INFN project (Iniziativa Specifica) SPIF.
I thank Pietro Colangelo and Fulvia De Fazio for their collaboration.

\end{document}